\documentstyle[aps,multicol,prb,epsf]{revtex}

\begin{document}

\title{Orbital character of O 2$p$ unoccupied states near the Fermi
  level in CrO$_2$}

\author{C. B. Stagarescu,\cite{corresp_author} X. Su, and D. E. Eastman}
\address{James Franck Institute, The University of Chicago, 5640 South
  Ellis Avenue, Chicago, Illinois 60637}

\author{K. N. Altmann and F. J. Himpsel}
\address{Department of Physics, University of Wisconsin Madison, 
  1150 University Avenue, Madison, Wisconsin 53706-1390}

\author{A. Gupta}
\address{IBM T. J. Watson Research Center, Yorktown Heights, 
  New York 10598}

\date{\today; submitted for publication, October 12, 1999}

\maketitle
\begin{abstract}
The orbital character, orientation, and magnetic polarization of the
O 2$p$ unoccupied states near the Fermi level ($E_F$) in CrO$_2$ was
determined using polarization-dependent X-ray absorption spectroscopy
(XAS) and X-ray magnetic circular dichroism (XMCD) from high-quality, 
single-crystal films. 
A sharp peak observed just above $E_F$ is excited only by
the electric field vector ($\bf E$) normal to the tetragonal
$c$-axis, characteristic of a narrow band ($\approx$ 0.7 eV bandwidth)
constituted from O 2$p$ orbitals perpendicular to $c$ (O 2$p_y$) 
hybridized with Cr 3$d_{xz-yz}$ $t_{2g}$ states. 
By comparison with band-structure and configuration-interaction (CI)
cluster calculations our results support a model of CrO$_2$ as a
half-metallic ferromagnet with large exchange-splitting energy 
($\Delta_{exch-split}$ $\approx$ 3.0 eV) and substantial correlation 
effects.
\end{abstract}
\begin{multicols}{2}
\indent Half-metallic ferromagnets are materials exhibiting
metallic character for the majority-spin electrons but a
semiconducting gap for the minority-spin electrons.\cite{groot83}
Correspondingly, one expects complete spin polarization at the Fermi
level, making this class of materials ideal for spin-polarized
emitters to be used in magnetic tunneling applications such as
magnetic random access memory\cite{gallagher97,daughton97} or
spin-polarized scanning tunneling microscopy.\cite{wiesendanger90}
CrO$_2$ is a ferromagnetic metal with a Curie temperature of $T_C$
$\approx$ 390 K.  Recently, it has generated substantial interest
\cite{mass,korotin98,lewis97,schutz97,soulen98,tsujioka97}
because its band structure has been predicted to
be half-metallic.\cite{korotin98,lewis97,schwarz86}
Point contact spectroscopy measurements of a spin polarization of about
90\% have been reported recently,\cite{soulen98} providing experimental
indication of its half-metallic character.\\
\indent 
Recent advances\cite{gupta} have made possible the growth 
of high-quality, epitaxial thin films of CrO$_2$ by 
chemical vapor deposition. 
Samples used in our experiment were as-prepared 4000 \AA~thick CrO$_2$ 
films grown on (100) TiO$_2$ substrates.
X-ray diffraction indicated that the films were single phase and the 
CrO$_2$ in-plane $<$100$>$ and $<$001$>$ ($a_\parallel$ and $c$) axes 
were aligned with the respective axes of the TiO$_2$ substrate. 
The third axis, crystallographically  equivalent to $a_\parallel$ and
normal to the surface of the films, will be denoted as $a_\perp$. 
The films exhibited a sharp ferromagnetic transition with a Curie
temperature of about 393 K and a strong in-plane magnetic anisotropy 
with $c$ and $a_\parallel$ being the easy-axis and hard-axis directions,
respectively. 
Cr$_2$O$_3$ spectra were taken from high-purity pressed-powder 
samples.\cite{cr2o3}
Our experiments were performed at the high-resolution Hermon 
beamline\cite{hermon} of the University of Wisconsin-Madison 
Synchrotron Radiation Center.
The XAS spectra were taken with a resolution of about 150 meV 
at 550 eV by measuring the total electron yield.
The degree of linear polarization was 85-90\%.
The XMCD measurements were taken with circularly polarized light
($\approx$ 80\%) from CrO$_2$ samples in switching remnant 
magnetization along the easy-axis which was positioned horizontally
and situated at 45$^{\circ}$ with respect to the direction of the
incoming radiation.\\
\indent To elucidate the spatial orientation and orbital character of 
the O 2$p$ empty states we have employed polarization-dependent XAS
measurements with linearly polarized light.
The results are shown in Fig. 1(a).
The CrO$_2$ spectrum displays a sharp peak (1) at $\approx$ 529.2 eV, 
followed by two other features (2) and (3) at energies of 2.3 and 
3.4 eV higher.
The sharp peak (1) exhibits a large intensity variation of 
$\approx$ 90\%, with a maximum for the geometry with the 
electric field vector of the light {$\bf E$} arranged 
along $a_\parallel$ (normal to the $c$-axis).
As shown further in Fig. 1(b), the $\phi$ dependence
of (1) follows accurately a cos$^{2}(\phi)$ curve. 
Our fit to the data, together with the degree of linear polarization 
of 85-90\%, gives an orientation of peak (1) 
perpendicular to the $c$-axis of more than 95\%.\cite{ellipt} 
The azimuthal dependence of the higher lying features (2) and
(3) is considerably smaller exhibiting an excursion of about 
28\% between the 0$^\circ$ and 90$^\circ$ orientations. 
At the same time its sign is opposite to that of (1) which indicates a 
dominant contribution from O 2$p$ states parallel to the $c$-axis.\\
\indent The O $K$ XAS polarization dependence has its origin
in the orientation of the O 2$p$ - Cr 3$d$ hybrid states with 
respect to the crystalline axes. 
The building blocks of the CrO$_2$'s rutile structure are CrO$_6$ 
octahedra arranged along and slightly elongated in the direction of the 
tetragonal $c$ axis.  
The O 2$p$ - Cr 3$d$ hybridization mechanism can be easily 
understood in a Goodenough molecular-orbital model electronic
structure scheme.\cite{goodenough}
The ligand crystal-field splits the Cr 3$d$ manifold into three 
$t_{2g}$ and two $e_g$ states. 
Due to the distinction between the $a_\perp$($a_\parallel$) and the 
$c$-axes in the tetragonal rutile structure,
\begin{figure}
\epsfxsize=70mm
\vspace{1mm}
\centerline{\epsffile{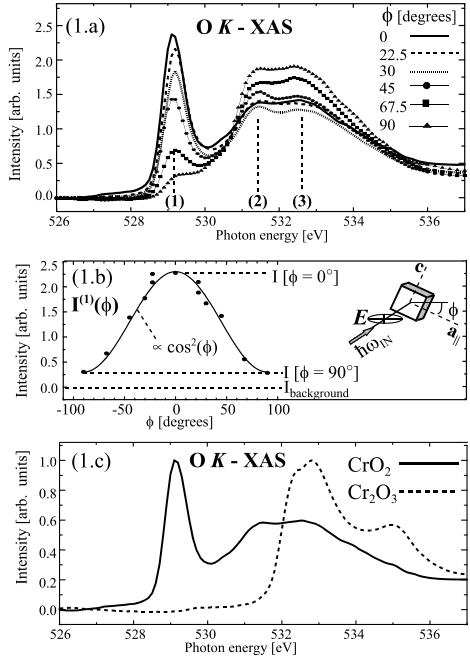}}
\vspace{1mm}
\narrowtext 
\caption {(a). O $K$ XAS of CrO$_2$ taken in a geometry with fixed polar
angle between the $c$ axis and the photon direction
($\theta$=90$^{\circ}$) and variable azimuthal angle ($\phi$) between
the photon direction and $a_\parallel$. 
(b). Intensity variation of the peak (1) with $\phi$. 
The experimental geometry is shown in the inset.
(c). Comparison of the O $K$ XAS of CrO$_2$ and Cr$_2$O$_3$.
}
\label{fig1}
\end{figure}
\noindent
the $t_{2g}$ group of Cr 
3$d$ states undergoes a further, small degeneracy splitting into 
an orbital normal to the $c$-axis ($d_{xz-yz}$) and two other 
having non-zero $c$-axis projections ($d_{xz+yz}$ and $d_{xy}$).
Oxygen atoms are situated in a trigonal environment forming planar
CrO$_3$ clusters with the three closest Cr atoms, with the planes of 
these CrO$_3$ clusters running parallel to the $c$ axis and oriented 
at 45$^\circ$ with respect to the $a_\parallel$ and $a_\perp$ directions. 
This trigonal ligand environment of oxygen leads to two $sp^2$
combinations (O 2$s$ and 2$p_{x,y}$) with non-zero $c$-axis
projections situated in the planes of the CrO$_3$ clusters and 
one remaining $p$ state (O 2$p_y$) that resides in the
$a_\parallel$-$a_\perp$ planes.
The Cr $d_{xz-yz}$ states are therefore
properly oriented to lead to $\pi$-type bonding with the O 2$p$ 
orbital which is perpendicular to the $c$-axis, i.e., O 2$p_y$.
Thus, the preceding analysis of the polarization dependence 
allows us to clearly identify peak (1) with empty states normal to the 
tetragonal $c$-axis formed by O 2$p_y$ hybridized with Cr 
3$d_{xz-yz}$ $t_{2g}$.
This result is confirmed by a recent band-structure calculation
of CrO$_2$ which finds that in the first $\approx$ 2 eV above
$E_F$ the empty O 2$p$ states are 100\% oriented perpendicular to
the $c$-axis.\cite{korotin_new}\\
\indent Even under normal
pressure/temperature conditions CrO$_2$ (Cr$^{4+}$) undergoes a 
chemical reduction to Cr$_2$O$_3$ (Cr$^{3+}$) which is the stable 
oxidation state of Cr.
For our samples, we estimated by tunneling measurements a 
Cr$_2$O$_3$ surface layer thickness of about 10~\AA.
We have addressed the issue of spectral contamination effects of the 
native insulating surface layer by measuring O $K$ XAS from 
Cr$_2$O$_3$ on the same experimental setup.
The results are shown in Fig. 1(c).
In contrast to CrO$_2$, the Cr$_2$O$_3$ O $K$ XAS 
spectrum displays significant intensity only above 532 eV 
having the highest intensity at about 532.8 eV.
Therefore we can conclude that the energy region from 529 to about 
532 eV in our CrO$_2$  O $K$ XAS does not contain any spectral 
interference from the surface layer of Cr$_2$O$_3$.\\
\indent Our CrO$_2$ O $K$ XAS spectra (peak (1) at 529.2 eV) have been
taken with a resolution of 150 meV and an accuracy of the absolute 
photon energy scale of about $\pm$ 100 meV.
Previously reported O $K$ absorption energies for peak (1)
taken from powder CrO$_2$ are of 527.0 eV\cite{tsujioka97}, 
530.0 eV\cite{langmuir}, and $\approx$ 530.5 eV\cite{schutz97}.
Our experience with measuring the CrO$_2$ and Cr$_2$O$_3$ absorption
spectra suggests that these differences are due to monochromator
absolute energy calibration errors rather than to sample
differences.
The previously reported CrO$_2$ O 1$s$ x-ray photoemission (XPS)
binding energy\cite{ikemoto76}, $E_B$ = 529.3 eV, is 
0.1 eV above the absorption peak (1).
Taking into account the limited accuracies of the two measurements,
an energy difference of 0.7$\pm$0.3 eV remains between the O 1$s$ 
XPS binding energy and the measured absorption edge at $\approx$ 528.6 eV,
where we assume a 0.2 eV accuracy of the XPS measured binding energy.\\
\indent 
The electron excited in the absorption process
interacts with its environment which
consists of the other valence electrons (correlation) 
and the positive charge of the core-hole 
(additional attractive potential).
To understand the effect of these interactions on the
absorption spectra we will compare our data with recent 
CI\cite{vanelp} and band structure calculations.\cite{korotin98,lewis97} 
The CI calculations start with a simplified 
model of a transition metal oxide as a MO$_6$ cluster
formed by a transition metal (M, d$^n$) ion situated in the octahedral 
ligand field of six oxygen ions (O$^{2-}$). 
Then, the valence electronic structure is described by   
allowing for $d-d$ Hubbard charge transfer ($d^{n}d^{n} \rightarrow
d^{n-1}d^{n+1}$), metal-ligand charge-transfer (e.g. $d^{n} \rightarrow 
d^{n+1}\underline{L}$ where $\underline{L}$= valence oxygen hole),
and metal-ligand hybridization which results in a ligand-hole
content in the ground-state. 
The oxygen $K$ absorption spectrum is obtained by calculating the 
differences in total energy (and the strength of the transitions)
between excited-states resulting from anihilation
of a valence ligand-hole by addition of one electron to a mixed metal-oxygen state 
(e.g. $d^{n+1}\underline{L} \rightarrow d^{n+1}$).
In Fig. 2 we compare the calculated CI oxygen XAS spectrum with the
experimental spectra of two isoelectronic 3$d^2$ oxides, 
CrO$_2$ (Cr$^{4+}$) and V$_2$O$_3$ (V$^{3+}$).\cite{horn} 
First, there is a good similarity between the O $K$ XAS 
spectra of the two 3$d^2$ oxides which indicates that,
due to strong $p-d$ hybridization,
the O $K$ edge absorption is mainly determined by the 
underlying 3$d$ electronic structure.
Second, a good correspondence is found between the three strongest
CI transitions and the three most prominent experimental
absorption features, which validates the choice of parameters
used in the calculation.
\begin{figure}
\epsfxsize=70mm
\vspace{1mm}
\centerline{ \epsffile{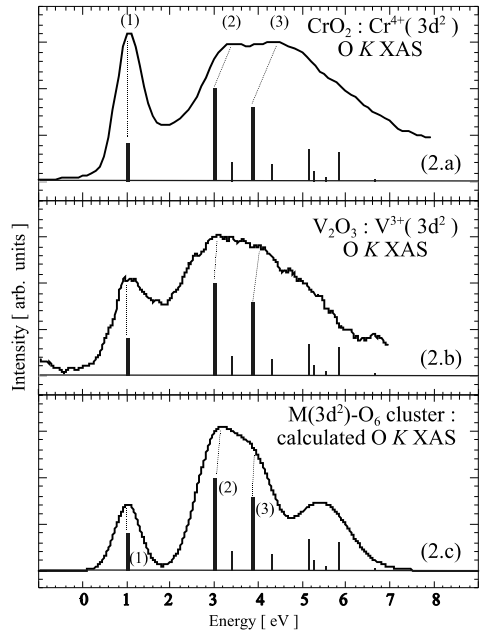}} 
\vspace{1mm}
\narrowtext
\caption {Comparison between experimental and CI calculated
O $K$ absorption for 3$d^2$ oxides.
(a). CrO$_2$ O $K$ XAS (spatially averaged, counting twice the doubly
degenerate $a$-axis ($a_\parallel$,  $a_\perp$) and once the $c$-axis). 
(b). V$_2$O$_3$ O $K$ absorption spectrum (representative for
V$_2$O$_3$ in the metallic state).\cite{horn}
(c). CI calculation.\cite{vanelp}
The common energy scale is that of the CI spectrum,
the first strong measured absorption features being aligned to the
first CI excited state.
For clarity the three strongest final states are shown with thicker
lines and the discrete excitation spectrum is replotted in panels
2(a) and 2(b). 
The continuous curve in 2(c) was obtained from the discrete CI
spectrum by a Gaussian convolution that simulates the effect of bandwidth
broadening (0.7 eV at (1) and 1.0 eV for higher energies).
}
\label{fig2}
\end{figure}
\noindent
Finally, two contributions can be proposed as the origin of the
small energy difference ($\approx$ 0.4 eV) observed 
for the CrO$_2$ features (2) and (3) (Fig. 2(a)).
First, a more accurate description of CrO$_2$ would have
to incorporate the deviation of the ligand-field from octahedral 
symmetry due to the tetragonal crystal structure distortion.
Second, the attractive core-hole potential 
(not included in the CI calculation) can, in principle, be expected to have
a stronger effect (increased downward energy shift) on the narrow, 
more localized $t_2g$ peak (1) than on the broader features (2) and (3).\\
\indent
The two band-structure calculations used in this study\cite{korotin98,lewis97} 
are both based on the  local-spin-density approximation of the 
density-functional theory (LSDA). 
As far as the low-energy excitation spectrum is concerned
(e.g. transport properties) both calculations agree in
finding CrO$_2$ to be a half-metal with a low density 
of fully spin-polarized states at  $E_F$.
However the two calculations differ importantly in the values obtained 
for the exchange-splitting energy responsible for ferromagnetism\cite{explain_exchsplit} 
($\Delta_{exch-split}$$\approx$3.0 eV in Ref. 6, 
$\Delta_{exch-split}$$\approx$1.8 eV in Ref. 7)
and in the treatment of the on-site electron-electron correlation.
In the LSDA approach of Lewis $et~al.$\cite{lewis97} 
the effects of electron-electron correlation are included in an 
averaged, mean-field way.  
In contrast, Korotin $et~al.$\cite{korotin98} use a LSDA scheme 
modified by the explicit inclusion of a potential correction\cite{pickett} (U=3 eV) 
in order to account for the Coulomb interaction between localized $d$ 
electrons. \\
\indent
We find a good similarity between the O $K$ XAS and the envelope of 
the O 2$p$ unoccupied DOS obtained by Korotin
$et~al.$, as can be observed by comparing spectra in panels (a) and
(b) of Fig. 3.
In particular, the amount of orbital overlap included in the 
calculation results in a band broadening consistent with the measured
width ($\approx$ 0.7 eV) of the O 2$p_y$ - Cr 3$d_{xz-yz}$ $t_{2g}$
peak (1) (Fig. 3(a)).
The two other O $K$ XAS features, (2) and (3) at higher energies,
correspond to local DOS maxima that have a mixed O 2$p$ - Cr 3$d$
$e_g$ spin-up and O 2$p$ - Cr 3$d$ $t_{2g}$ spin-down character. 
It becomes clear when looking at the theoretical spin-up/spin-down
projected DOS in Fig. 3(c) and 3(d) that the essential parameter
controlling the good agreement with the LSDA+U calculation
(i.e. the appearance of a broad minimum in the empty DOS at 
$\approx$1.5 eV above $E_F$ and of significant local maxima (2) and
(3) at higher energies, as observed experimentally) is the
large upward shift of the spin-down DOS ($\Delta_{exch-split}$$\approx$3.0 eV).
Our data are not well reproduced by the LSDA calculation\cite{lewis97} 
which has $\Delta_{exch-split}$$\approx$1.8 eV.
This smaller exchange-splitting leads directly to a spin-down DOS
filling in the region of the broad experimental minimum ($\approx$1.5
eV above $E_F$) and to the presence of the other (higher energy)
significant DOS features (mainly due to spin-down $t_{2g}$ states) 
shifted closer to $E_F$, as seen in Fig. 3(e).
An experimental $\Delta_{exch-split}$ can be determined
in a ''rigid band'' approximation by slightly adjusting the 
(LSDA+U) exchange-splitting energy shift to fit our data (features (2) and
(3)).
This ``DOS-fit'' yields $\Delta_{exch-split}$ $\approx$ 3.2 eV.
Given the uncertainties in this fitting\cite{correction}
and experimental accuracies, we conclude that our data 
indicate a $\Delta_{exch-split}$ $\approx$ 3.0 eV. 
This is consistent with the LSDA+U calculation,\cite{korotin98}
in which a large $\Delta_{exch-split}$ was obtained as an explicit 
consequence of including a considerable correlation energy (U=3eV). 
Thus, our experimental results supports the conclusion of the LSDA+U
work\cite{korotin98} that on-site correlation effects 
beyond the average LSDA value are present CrO$_2$.
Interestingly, we also observe that the correlation energy parameter U
is consistent (same origin) for the LSDA+U calculation\cite{korotin98} 
(obtained by the local-orbital expansion method described by 
Pickett $et~al.$\cite{pickett}) and the CI calculation.\cite{vanelp,vanelp_mno}\\
\indent 
Finally, as the first empty oxygen states have been experimentally determined
to be of O 2$p_y$ origin, a substantial magnetic orbital polarization
of these states is expected since their Cr 3$d_{xz-yz}$ $t_{2g}$ 
counterparts are 
fully spin-polarized (Fig. 3(d)).
Indeed, we have measured
a clear O $K$ XMCD  signal (not shown here), similar to 
\begin{figure}
\epsfxsize=70mm
\vspace{1mm}
\centerline{ \epsffile{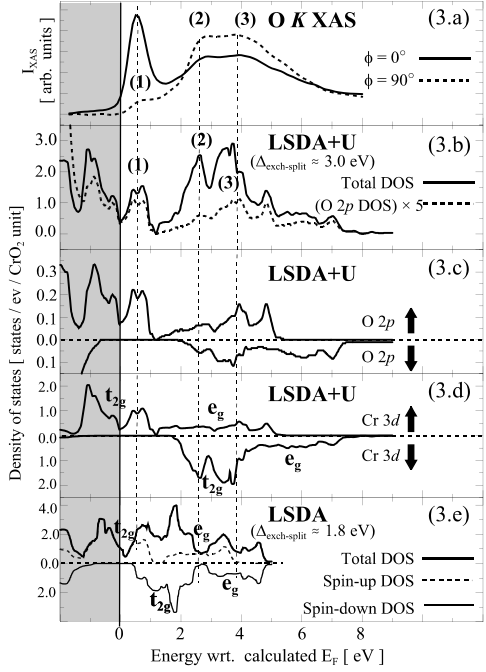}} 
\vspace{1mm}
\narrowtext
\caption {Comparison of the CrO$_2$ O $K$ XAS with calculated DOS. 
The theoretical DOS are plotted against their calculated binding energy scales. 
The XAS peak (1) is aligned to the isolated set of empty states 
(1) in the calculated DOS.
(a). CrO$_2$ O $K$ XAS for two azimuthal angles $\phi$ = 0$^\circ$ and
90$^\circ$.
(b). Total DOS and O 2$p$ partial DOS from the LSDA+U calculation of
Korotin $et~al.$\cite{korotin98}
(c, d). O 2$p$ and Cr 3$d$ spin resolved DOS from the same LSDA+U work.\cite{korotin98} 
(e). Total, spin-up, and spin-down DOS from the LSDA calculation of Lewis
$et~al.$\cite{lewis97}
}
\label{fig3}
\end{figure}
\noindent
that obtained previously by Attenkofer $et~al.$\cite{schutz97} from polycrystalline
CrO$_2$.\\
\indent
In conclusion, we used polarization-dependent XAS to determine
the orbital character of the unoccupied O 2$p$ states just above $E_F$
in CrO$_2$ which are found to be O 2$p_y$ states hybridized with Cr
3$d_{xz-yz}$ $t_{2g}$, with a bandwidth of $\approx$ 0.7 eV.
A large experimental exchange-splitting energy of $\Delta_{exch-split}$
$\approx$ 3.0 eV was estimated by comparison between our data and
band-structure calculations.
Our results support a model of CrO$_2$ as a 
half-metallic ferromagnet with substantial correlation effects.\\\\
\indent 
The help of X. W. Li in growth of the sample and of
M. Bissen and R. Hansen in setting the experiment 
is gratefully acknowledged.
The authors thank M. A. Korotin for communicating his results
prior to publication. One of the authors (C. B. S.) thanks 
G. A. Sawatzky and D. I. Khomskii for useful discussions.
This work was supported in part by the DOE under 
Contract W-31-109-ENG-38. The Synchrotron Radiation
Center, UW-Madison, is supported by the NSF under Award No.
DMR-9531009.

\end{multicols}

\begin{thebibliography}{99}
\bibitem{explain_exchsplit}
 The LSDA work (Ref. 7) gives $\Delta_{exch-split}$=1.8 eV. 
 For the LSDA+U calculation, we have estimated
 $\Delta_{exch-split}$ $\approx$ 3.0 eV by measuring the energy between the 
 ``centers of mass'' of the distinct Cr 3$d$ $t_{2g}$
 spin-up/spin-down DOS.
\bibitem{correction}
  If the apparent increase in the (1)-(2) energy separation 
  (Fig. 2) is confirmed 
  (e.g by comparison with a more accurate CI calculation)
  to be a core-hole effect, 
  $\Delta_{exch-split}$ would need to be corrected to $\approx$ 2.8 eV.
\end{thebibliography}
\end{document}